\documentclass[journal]{IEEEtran}
\usepackage{amsmath,amsfonts}
\usepackage{mathtools} 
\usepackage[caption=false,font=scriptsize,labelfont=rm,textfont=rm,subrefformat=parens]{subfig}
\usepackage{algorithmic}
\usepackage{algorithm}
\usepackage{array}
\usepackage{textcomp}
\usepackage{pifont}
\usepackage{stfloats}
\usepackage{url}
\usepackage{verbatim}
\usepackage{graphicx}
\usepackage{float}
\usepackage{cite}
\usepackage{booktabs}
\usepackage{multirow}
\usepackage{bigstrut}
\usepackage{colortbl}
\usepackage{afterpage}
\usepackage{xcolor}
\usepackage{soul}
\graphicspath{{Figures/PDF/}{Figures/PNG/}}
\usepackage[colorlinks=true,linkcolor=blue,citecolor=blue,urlcolor=blue,]{hyperref}
\usepackage{orcidlink} %调包

\DeclareSubrefFormat{myparens}{#1(#2)}
\usepackage{xcolor}
\usepackage{xpatch}

\makeatletter
\ExplSyntaxOn
\cs_new:Npn \bibColoredItems #1#2
  {
    \clist_map_inline:nn {#2} { \cs_new:cpn {bib@colored@##1} {#1} } 
  }
\ExplSyntaxOff

\newcommand\bib@setcolor[1]{%
  \ifcsname bib@colored@#1\endcsname
    \expanded{\noexpand\color{\csname bib@colored@#1\endcsname}}%
  \else
    \normalcolor
  \fi
}

\IfPackageLoadedTF{hyperref}{\@tempswatrue}{\@tempswafalse}
\if@tempswa
  \xpatchcmd\@bibitem {\H@item}{\bib@setcolor{#1}\H@item}{}{\PatchFailed}
  \xpatchcmd\@lbibitem{\H@item}{\bib@setcolor{#2}\H@item}{}{\PatchFailed}
\else
  \xpatchcmd\@bibitem {\item}  {\bib@setcolor{#1}\item}  {}{\PatchFailed}
  \xpatchcmd\@lbibitem{\item}  {\bib@setcolor{#2}\item}  {}{\PatchFailed}
\fi
\makeatother

\definecolor{revisioncolor}{HTML}{000000} %预定义高亮颜色revisioncolor，直接引用，统一修改。

\begin{document}

% \textsuperscript{\orcidlink{0009-0002-2830-0876}}

\title{Overcoming the Identity Mapping Problem in Self-Supervised Hyperspectral Anomaly Detection}

% \author{Yongchuan Cui\textsuperscript{\orcidlink{0009-0002-5179-6232}},~\IEEEmembership{Graduate~Student~Member,~IEEE,} 
\author{Yongchuan Cui, 
        Jinhe Zhang,
        Peng Liu,~\IEEEmembership{Senior~Member,~IEEE,}
        % Peng Liu\textsuperscript{\orcidlink{0000-0003-3292-8551}},~\IEEEmembership{Senior~Member,~IEEE,}
        Weijing Song, 
        and Yi Zeng
        % <-this % stops a space
\thanks{This research was partially supported by the National Natural Science Foundation of China (NSFC) under Grant 41971397, and Project E3E2181101, 2024YFF1307204 \textit{(Corresponding author: Peng Liu. Email: \mbox{liupeng202303@aircas.ac.cn}).}}
\thanks{Yongchuan Cui and Peng Liu are with Aerospace Information Research Institute, Chinese Academy of Sciences, Beijing, China, and School of Electronic, Electrical and Communication Engineering, University of Chinese Academy of Sciences, Beijing, China. Jinhe Zhang is with School of Geography and Information Engineering, China University of Geosciences (Wuhan), Wuhan, China. Weijing Song is with School of Computer Science, China University of Geosciences (Wuhan), Wuhan, China. Yi Zeng is with College of Information, Beijing Forestry University, Beijing, China.}
\thanks{Yongchuan Cui and Jinhe Zhang are co-first authors.}
}

\markboth{IEEE Transactions on Geoscience and Remote Sensing}%
{Shell \MakeLowercase{\textit{et al.}}: A Sample Article Using IEEEtran.cls for IEEE Journals}


\maketitle

\begin{abstract}
	The surge of deep learning has catalyzed considerable progress in self-supervised Hyperspectral Anomaly Detection (HAD). The core premise for self-supervised HAD is that anomalous pixels are inherently more challenging to reconstruct, resulting in larger errors compared to the background. However, owing to the powerful nonlinear fitting capabilities of neural networks, self-supervised models often suffer from the Identity Mapping Problem (IMP). The IMP manifests as a tendency for the model to overfit to the entire image, particularly with increasing network complexity or prolonged training iterations. Consequently, the whole image can be precisely reconstructed, and even the anomalous pixels exhibit imperceptible errors, making them difficult to detect. Despite the proposal of several models aimed at addressing the IMP-related issues, a unified descriptive framework and validation of solutions for IMP remain lacking. In this paper, we conduct an in-depth exploration to IMP, and summarize a unified framework that describes IMP from the perspective of network optimization, which encompasses three aspects: perturbation, reconstruction, and regularization. Correspondingly, we introduce three solutions: superpixel pooling and uppooling for perturbation, error-adaptive convolution for reconstruction, and online background pixel mining for regularization. With extensive experiments being conducted to validate the effectiveness, it is hoped that our work will provide valuable insights and inspire further research for self-supervised HAD. Code: \url{https://github.com/yc-cui/Super-AD}. 
\end{abstract}
\begin{IEEEkeywords}
    Hyperspectral anomaly detection, identity mapping, deep learning, self-supervised neural networks
    \end{IEEEkeywords}

\section{Introduction}
\label{sec:intro}

\begin{figure}[t]
  \centering
  {\includegraphics[width=1\linewidth]{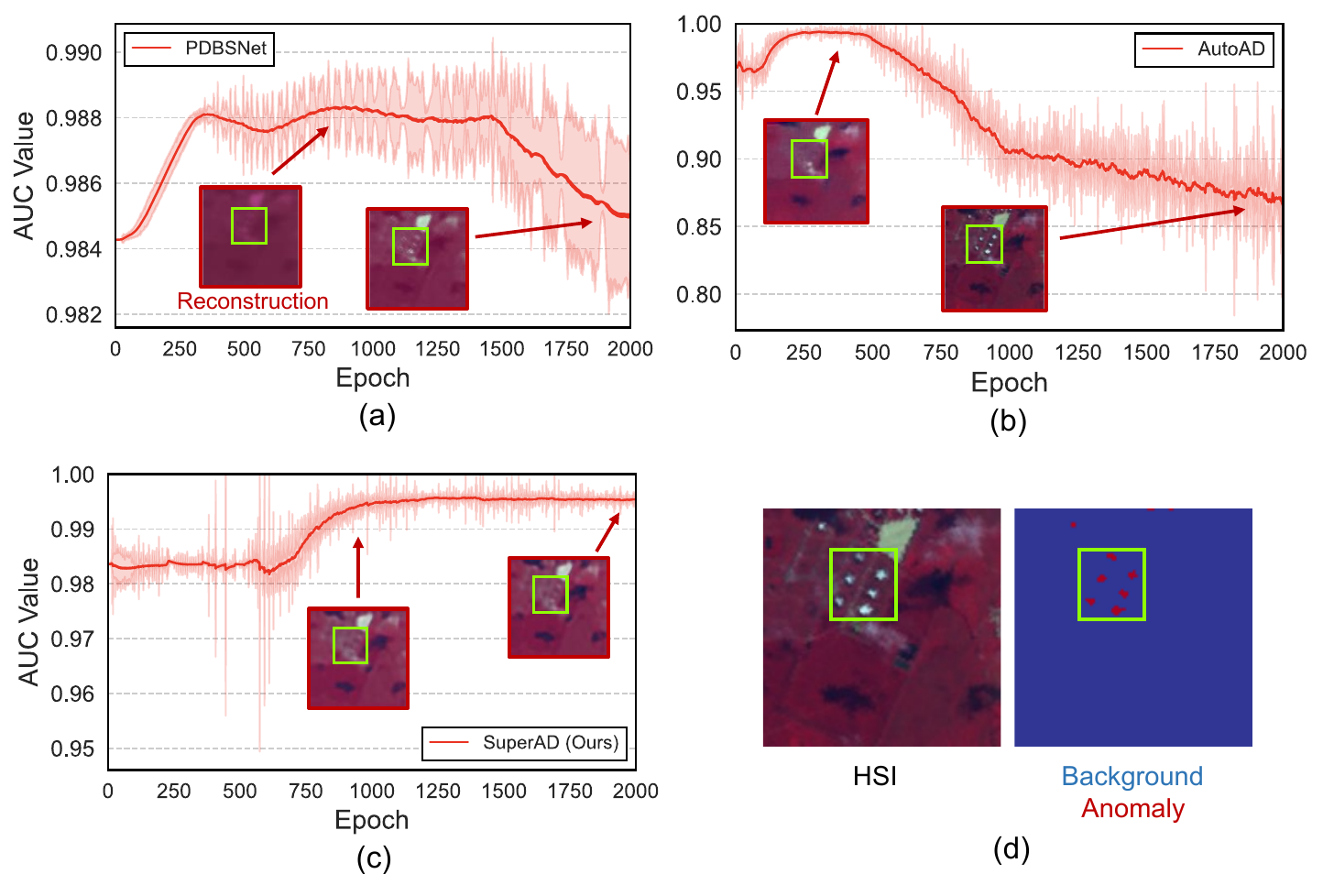}}
  \caption{Comparison of reconstruction process across different models: (a) PDBSNet~\cite{PDBSNet}, (b) AutoAD~\cite{AutoAD}, and (c) our proposed SuperAD model. (d) shows the hyperspectral image and the groudtruth detection map. The fluorescent green rectangular regions highlight critical areas of interest that demonstrate the identity mapping phenomenon.}
  \label{fig:intro-imp}
\end{figure}

Hyperspectral anomaly detection (HAD) aims to identify pixels or regions in a hyperspectral image that exhibit spectral signatures significantly different from surroundings~\cite{9205919,su2021hyperspectral,bioucas2013hyperspectral}. Traditional methods for HAD have relied heavily on statistical approaches, such as the Reed-Xiaoli (RX)~\cite{RXD,1386510} detectors, collaborative representation (CR)~\cite{CRD,8561244} and low-rank representation (LRR)~\cite{7322257,liu2010robust}. These methods, while effective in certain scenarios, often struggle with the complexity and variability of real-world hyperspectral data, leading to suboptimal detection performance~\cite{7322257}. Compared to traditional methods, the advent of parameterized neural networks for self-supervised learning~\cite{AutoAD,PDBSNet,BiGSeT,BockNet,SMCNet,DirectNet,wu2024transformer}, has emerged as a promising approach in HAD. The core premise is that the background, comprising the majority of the image, can be approximated well by the model, while anomalies, being spectrally distinct, cannot be accurately represented by the learned background model~\cite{7322257,BS3LNet}. However, self-supervised models in HAD face a significant challenge known as the Identity Mapping Problem (IMP), which has been extensively mentioned in~\cite{DirectNet,BiGSeT,PDBSNet,BockNet,BS3LNet, rs16163036}. The IMP arises from the powerful nonlinear fitting capabilities of deep neural networks, which can lead to overfitting to the entire image dataset. As the complexity of the network increases or the number of training iterations grows, these models tend to reconstruct both the background and anomalies with high fidelity, resulting in imperceptible errors for anomalous pixels~\cite{9715082}. Despite the introduction of various models attempting to tackle IMP-related issues, a comprehensive analytical framework and a unified validation of solutions for IMP in the context of self-supervised HAD are still missing, lacking a holistic view of the problem. In this paper, we aim to fill this gap by conducting an in-depth exploration to the IMP. Specifically, we propose a unified framework (Super-AD) that describes the IMP from the perspective of network optimization, encompassing three key aspects: perturbation, reconstruction, and regularization. Each aspect corresponds to a specific solution that we introduce. Through extensive experiments on various hyperspectral datasets, we validate the effectiveness of our proposed solutions and demonstrate how they collectively contribute to overcoming the IMP.

To better understand the IMP and demonstrate the effectiveness of our proposed approach, Fig.~\ref{fig:intro-imp} presents a comparative analysis of area under the receiver operating characteristic curve (AUC) performance across different models when being optimized. Fig.~\ref{fig:intro-imp}(a) and (b) illustrate the AUC trends of two state-of-the-art self-supervised models, PDBSNet~\cite{PDBSNet} and AutoAD~\cite{AutoAD}, respectively. As training iterations increase, both models exhibit a common pattern: after reaching their peak AUC values, their performance gradually declines. This phenomenon directly reflects the IMP, where the networks' powerful reconstruction capabilities lead to overfitting, causing them to reconstruct both background and anomalous pixels with high fidelity (as highlighted by the fluorescent green regions in the reconstruction images). In contrast, our proposed SuperAD model, shown in Fig.~\ref{fig:intro-imp}(c), maintains stable AUC performance even with increasing iterations, demonstrating its robustness against the IMP. The reconstruction results further validate that our model effectively preserves the distinction between background and anomalies, preventing the reconstruction of anomalous pixels.

The main contributions of this work can be summarized as follows:

\begin{itemize}
    \item We present the first comprehensive framework that systematically analyzes the identity mapping problem in self-supervised hyperspectral anomaly detection, providing a detailed theoretical foundation and practical insights into this critical issue.

    \item We propose three key strategies to address the IMP: (1) superpixel-based pooling and uppooling operations to enhance spatial-spectral feature representation, (2) error-adaptive convolution to dynamically adjust feature learning based on reconstruction errors, and (3) online background pixel mining to improve model robustness against anomalies. Each component has been thoroughly visualized to demonstrate its effectiveness, including superpixel pooling visualization, pixel utilization visualization, and background mining process visualization.

    \item Extensive experiments on multiple hyperspectral datasets demonstrate the effectiveness of our proposed solutions in mitigating the IMP and improving anomaly detection performance compared to state-of-the-art methods. Our work provides valuable insights and a solid foundation for future research in self-supervised hyperspectral anomaly detection, offering a unified framework that can be extended and adapted to various related applications.
\end{itemize}

\section{Related Work}\label{sec:related}

In recent years, one of the key technologies in remote sensing is hyperspectral imaging and associated anomaly detection (HAD). It is frequently employed in both military and civilian domains because of its feature. However, because of environmental light variations, interference like atmospheric scattering, the mixed pixel problem brought on by low spatial resolution, and the absence of anomaly sample labels~\cite{keshava2002spectral,shi2014incorporating,nishii1996enhancement}, HAD currently depends on the statistical characteristics of the data itself or deep features to distinguish the background from the anomaly. Model-based approaches~\cite{schweizer_efficient_2001} and deep learning-based methods~\cite{schweizer_efficient_2001,chang2007hyperspectral,li_learning_2024,Cui2024Semi} are the two main classes of HAD techniques.

\subsection{Traditional Methods}

Model-driven anomaly detection techniques, which primarily consist of two primary algorithmic frameworks, statistical modeling and representation learning—dominate classical HAD research~\cite{su2021hyperspectral}. Reed introduced statistical HAD with the Reed-Xiaoli (RX) approach, which makes the assumption that the background follows a multivariate Gaussian distribution and that Mahalanobis distance thresholding is used to identify the anomalies~\cite{RXD,su2021hyperspectral}. A number of variations have been created to improve robustness, such as Kernel-RX (KRX)~\cite{su2021hyperspectral,kwon2005kernel}, Local-RX (LRX)~\cite{su2021hyperspectral,matteoli2010local}, Segmented-RX~\cite{su2021hyperspectral,matteoli2010improved} and Weighted-RX~\cite{su2021hyperspectral,guo2014weighted}. However, modeling the background distribution with a Gaussian distribution~\cite{matteoli2010tutorial,yang_compressive_2015,manolakis2001statistics} is insufficient due to the intricacy of HSI. 

Over time, representation learning-based detection frameworks have emerged as a research hotspot in an effort to overcome the reliance of conventional HAD techniques on data distribution assumptions. Such methods extract essential features of the data by constructing adaptive representation models. Among them, the widely used ones are low rank representation (LRR)~\cite{gao2021using,zhuang2021hyperspectral,zhuang2020hyperspectral,yin2015laplacian}, collaborative representation (CR)~\cite{zhao2022hyperspectral} and sparse representation (SR)~\cite{zhuang2021fasthymix,li2020lowrank,zhuang2023crosstrack,du_beyond_2016}. In order to mine the spatial-spectral correlation of hyperspectral data from a global perspective, low-rank representation (LRR)~\cite{liu2010robust} models interpixel correlations based on global low-rank constraints, separating anomalies by residuals of the original image from the low-rank background. The collaborative representation detector (CRD)~\cite{zhang2014hyperspectral}, which is based on the a priori assumption that anomalies are hard to represent linearly by neighbors, uses linear combinations of spatial neighborhood pixels to represent the current pixel and detects anomalies using the representation residuals. Nevertheless, the model-driven representation learning approach still has substantial drawbacks~\cite{bioucas2013hyperspectral}: the model hyperparameters (\textit{e.g.,} sparsity, rank constraints) must be manually adjusted, and the parameter settings are highly correlated with the scene; it is also not generalizable, and it is challenging to transfer to other sensors or feature type data once the model has been trained for a particular image.

\begin{figure*}[htbp]
    \centering\includegraphics[width=1\linewidth]{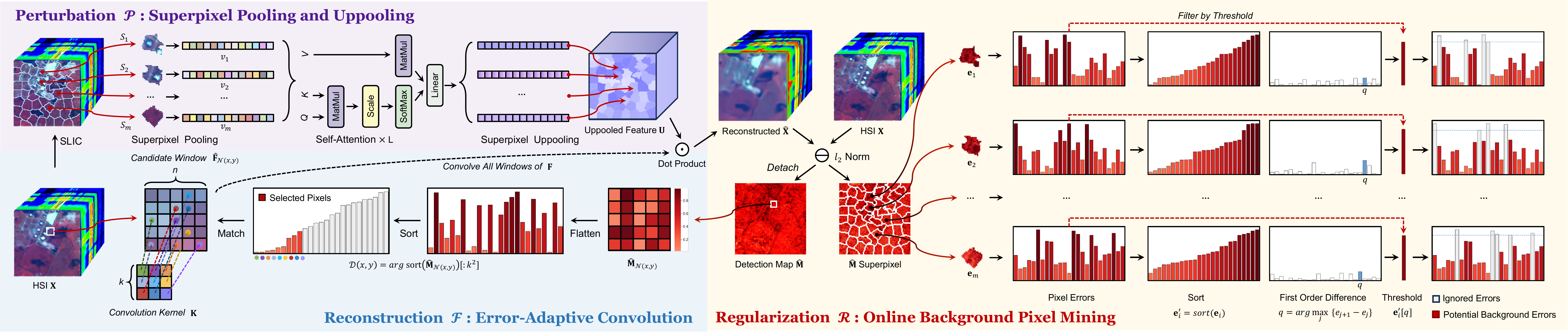}
    \caption{Our proposed unified framework for self-supervised HAD (Zoom in for better view), including the perturbation operation $\mathcal{P}(\cdot)$, the reconstruction function $\mathcal{F}(\cdot,\cdot;\theta)$, and the regularization term $\mathcal{R}(\cdot)$. For every part, we propose the corresponding solution to address the IMP, including the superpixel pooling and unpooling (SPP), error-adaptive convolution (AdaConv), and online background pixel mining (OBPM).} \label{fig:net}
  \end{figure*}

\subsection{Deep Learning-based Self-Supervised Methods}

Linear assumptions of traditional statistical methods and shallow feature extraction mechanisms make it difficult to fully explore the nonlinear associations and deep semantic information implied in HSI data. Deep learning excels in in nonlinear feature extraction, which has been extensively employed in HSI anomaly detection with demonstrated effectiveness~\cite{xu2022hyperspectral,xie2020autoencoder,Cui2025Pansharpening,sun2025decadedeeplearningremote}. Autoencoders (AEs)~\cite{zhang2016crop} and Generative Adversarial Networks (GANs)~\cite{jiang2020discriminative} form the main approaches of unsupervised deep learning.

On benchmark datasets, early research by Taghipour et al. showed that simple AE designs with $l_2$ reconstruction loss minimization achieve comparable performance, follow-up studies have shown that deepening the network reduce reconstruction error gaps between anomalous and background pixels, which is explained by the network's tendency to learn uncommon anomaly patterns thanks to its enhanced modeling capacity~\cite{taghipour2019unsupervised,wang2023digging,hartley2004minimization,Cui2024Reconstruction}. The emergence of GANs has provided a new approach to HAD: realistic backgrounds are generated using a training generator, and the discriminator recognizes the anomalies. Wang et al.~\cite{wang2023frequency} used this technique which demonstrated superior separability to achieve a significant improvement in anomaly separability compared to a self-encoder baseline~\cite{goodfellow2014generative,arjovsky2017wasserstein,xu2022hyperspectral,abuhani2025generative,Cui2024Pixel}. In GAN-based techniques, a discriminator locates the anomalies by detecting differences, while a generator is trained to create realistic background pixels. By integrating GAN adversarial training with AE reconstruction, AAE~\cite{xie2020autoencoder} limits the latent space distribution. HAD has recently seen major improvements with the advent of Transformer architectures. For instance, Transformers like SpectralFormer~\cite{wu2024transformer} represent global spectrum dependencies using a self-attentive method, achieving top results anomaly detection capabilities on benchmark hyperspectral datasets. However, the computational cost of these methods is greatly increased by their effectiveness~\cite{liu2024msnet}. This is why masked autoencoding techniques~\cite{vaswani2017attention} randomly exclude most spectral bands during training in an effort to strike a balance between performance and efficiency. Exploiting inter-band correlations has proven to have significant limitations in real-world deployments, despite its effectiveness in controlled trials~\cite{zhang2023spectrum}, particularly when working with data from innovative sensor systems.

Despite the recent advancements in HAD, identity mapping problem (IMP) related flaws are still present in current approaches. The IMP stems from the network's memorize anomaly patterns. First, over-parameterization is directly related to IMP susceptibility, with over-parameterized networks showing higher vulnerability~\cite{mou2017deep,luo2022frequency}. Second, IMP happens regardless of architectural changes, and the global reconstruction aim basically pushes the network in the direction of identity mapping. Third, anomalies can be accurately reconstructed from surrounding band information thanks to strong spectral correlation~\cite{li2024superpixel}. Let $\mathcal{F}$ be the neural network and $X$ be the input HSI of a self-supervised model (\textit{e.g.,} AE). When $\mathcal{F}(\mathbf{X}) \approx \mathbf{X}$ holds for every pixel, including anomalies, IMP takes place. The reconstruction error $\|\mathbf{X} - \mathcal{F}(\mathbf{X})\|$, which goes to zero and loses discriminative qualities, is what anomaly detection depends on. In this instance, the model's ability to distinguish normal and abnormal pixels is compromised, which lowers the efficacy of anomaly detection. The error of abnormal pixels that appear as unnoticeable abnormal pixels increases in HAD as the network complexity~\cite{chang2020effective} or the number of training iterations and the accurate reconstruction of anomalies grow. The more complex or well-trained the model, the more likely it is to learn the identity mapping ($\hat{X} \approx X$) that includes anomalies and background, rather than extracting discriminative features.

\section{Methodology}\label{Methodology}

\subsection{Unified Perspective for Self-supervised HAD}
% \vspace{-1mm}

In the context of self-supervised HAD, we address the IMP by proposing a unified framework that encompasses perturbation, reconstruction, and regularization. Given a hyperspectral image $\mathbf{X}\in \mathbb{R}^{h\times w \times c}$ containing anomalous pixels, we utilize a neural network $\mathcal{F}$ parameterized by $\theta$ to reconstruct the image. We summarize the optimization process of a theoretically well-performing neural network using the following unified formulation:
% \vspace{-3mm}
\begin{equation}
    \hat{\theta}  = \arg \min_{\theta}\mathcal{L} \left(\mathcal{F}\left(\mathcal{P}\left(\mathrm{\mathbf{X}}\right), \hat{\mathrm{\mathbf{M}}};\theta \right) , \mathrm{\mathbf{X}}\right) + \lambda \mathcal{R}(\hat{\mathrm{\mathbf{M}}}),
\end{equation}
where $\mathcal{L}(\cdot,\cdot)$ denotes the reconstruction loss (commonly $l_1$ or $l_2$), and $\hat{\mathbf{M}}$ is the estimated anomaly map from the previous iteration. The critical components of gaining insights into IMP include the data perturbation operation $\mathcal{P}(\cdot)$, the guided reconstruction function $\mathcal{F}(\cdot,\cdot;\theta)$, and the regularization term $\mathcal{R}(\cdot)$. $\lambda$ balances the contribution of the regularization. In the following, we will introduce the design of each component in detail. We argue that the recently proposed HAD methods related to the IMP can be incorporated into our framework.

\subsubsection{Perturbation}

The data perturbation operation $\mathcal{P}(\cdot)$ is designed to perturb the spectral information. Applying perturbations to obscure the information of anomalous spectra before they can influence the network's reconstruction process is a straightforward strategy to mitigate the IMP. The random masking strategy in SMCNet~\cite{SMCNet} and AETNet~\cite{AETNet}, the use of noise in AutoAD~\cite{AutoAD} and BSDM~\cite{BSDM}, \textit{etc.}, are specific perturbation instances. The blind spot network series, BS3LNet~\cite{BS3LNet}, BockNet~\cite{BockNet}, PDBSNet~\cite{PDBSNet}, DirectNet~\cite{DirectNet}, \textit{etc.}, can actually be considered a form of perturbation, which can be viewed as applying a mask to the central pixel during convolution.

\subsubsection{Reconstruction}

The reconstruction function $\mathcal{F}(\cdot,\cdot;\theta)$ leverages the estimated anomaly map $\hat{\mathbf{M}} \in \mathbb{R}^{h \times w}$ from the previous iteration to guide the forward process of the network. Given the premise that anomalous spectra are difficult to reconstruct, the reconstruction error from the previous iteration can be treated as a confidence measure for the anomaly map. By designing a suitable weighting function for $\hat{\mathrm{\mathbf{M}}}$, we can enhance the error associated with anomalous spectra while diminishing the error related to the background, leading to a more accurate anomaly map in subsequent iterations and thus mitigating the IMP. For instance, BiGSeT~\cite{BiGSeT} and MSNet~\cite{MSNet} utilized the dot product to modify the reconstruction results. AutoAD~\cite{AutoAD}, DeepLR~\cite{DeepLR}, and S2DWMTrans~\cite{S2DWMTrans} employed adaptive weights to alter the gradients during the backpropagation.

\subsubsection{Regularization}

This term imposes constraints on the estimated anomaly map to prevent the IMP. The weight coefficient $\lambda$ balances the contributions of the reconstruction loss and the regularization term. In the optimization process of the neural network, this term is typically formulated as a loss function that imposes additional constraints on the anomaly map. For instance, BiGSeT~\cite{BiGSeT} and MSNet~\cite{MSNet} applied the second-order Laplacian of Gaussian (LoG) operator to suppress anomalies. DeepLR~\cite{DeepLR} and RSAAE~\cite{RSAAE} applied a low-rank regularized loss to constrain the network to approximate the low-rank background. However, a common challenge in existing methods is the difficulty in determining the balance coefficient $\lambda$ between reconstruction and regularization.

Although each part presents various methods, their limited consideration of the reconstruction process from a holistic perspective of network optimization results in constrained performance. In this paper, we meticulously designed these three key aspects, and experiments prove that our approach can achieve optimal results (see Fig.~\ref{fig:net}).

\subsection{Design of the Perturbation Operation $\mathcal{P}$}

Masking~\cite{SMCNet,AETNet,BS3LNet,BockNet,PDBSNet,DirectNet} and noise~\cite{AutoAD,BSDM} cannot ensure the total elimination of anomalous spectra before sent into the network. To this end, we propose a new perturbation strategy, \textit{i.e.,} superpixel pooling and unpooling (dubbed as SPP). Specifically, we first use Simple Linear Iterative Clustering (SLIC)~\cite{SLIC} to segment the hyperspectral image into superpixels, and then apply average pooling to each region block to retain the average feature information. Since anomalies occupy a small proportion, they are easily wrapped in pixel blocks surrounded by the background. Due to the average pooling strategy, the block information will contain mostly background spectra while ignoring the anomalous spectra, which prevents the anomalous spectra from being reconstructed, thereby mitigating the IMP. Meanwhile, for the extracted all blocks, we use the self-attention mechanism~\cite{vaswani2017attention,ViT} to perform spectral reconstruction, learning the relationship between the blocks. Finally, all blocks will perform uppooling to revert to original size. Compared to masking~\cite{SMCNet,AETNet,BS3LNet,BockNet,PDBSNet,DirectNet} and noise~\cite{AutoAD,BSDM} strategies, SPP effectively encapsulates anomalous pixels within background-dominated blocks, thereby preventing their influence on the reconstruction process.

Formally, given a hyperspectral image $\mathrm{\mathbf{X}} \in \mathbb{R}^{h\times w \times c}$, $\operatorname{SPP}(\mathrm{\mathbf{X}})$ can be described as follows. Firstly, obtaining a series of superpixel blocks using the SLIC~\cite{SLIC} algorithm,
\begin{equation}
    \mathcal{S}=\operatorname{SLIC}(\mathrm{\mathbf{X}}),
\end{equation}
where $\mathcal{S}=\{S_1, S_2, \cdots, S_m\}$, $S_i$ represents the $i$-th superpixel. Then, we apply average pooling to each superpixel to obtain the feature vectors $\mathcal{V} = \{v_1, v_2, \cdots, v_m\}$. The pooling process can be expressed as,
\begin{equation}
    v_i=\frac{1}{|S_i|}\sum_{p\in S_i}\mathrm{\mathbf{F}}_p,
\end{equation}
where $\mathrm{\mathbf{F}}_p$ denotes the feature vector of pixel $p$ used in superpixel pooling and $\left|\cdot\right|$ is the cadinality of set (\# of pixels). After forward the self-attention~\cite{vaswani2017attention,ViT}, the feature vector will be restored to its original shape through uppooling,
\begin{equation}
    \mathrm{\mathbf{U}}(x,y)=\sum_{v_i \in \mathcal{V}} v_i \cdot 1_{p_{xy} \in S_i},
\end{equation}
where $\mathrm{\mathbf{U}}\in \mathbb{R}^{h\times w \times c}$ is the uppooled feature. $1_{p_{xy} \in S_i}$ is an indicator function that is $1$ if $p$ in $(x, y)$ belongs to $S_i$, and $0$ otherwise.

\subsection{Design of the Reconstruction Function $\mathcal{F}$}

Commonly, existing designs directly use the estimated anomaly map as a weight~\cite{AutoAD,BiGSeT,MSNet,DeepLR,S2DWMTrans}, which still allow anomalous pixels to affect the reconstruction process. In contrast, we propose a novel guided reconstruction mechanism termed error-adaptive convolution (dubbed as AdaConv), which maximizes the non-utilization of anomalies. AdaConv performs dynamic convolution only on pixels that are most likely to be non-anomalous based on anomaly probability from the previous iteration. 

Specifically, given a coordinate $(x,y)$, we get the indices of all elements of a candidate window of size $n\times n$, 
\begin{equation}
    \begin{split}
    \mathcal{N}(x,y) = \{(i,j) \mid & i \in [x - \frac{n-1}{2}, x + \frac{n-1}{2}], \\
    & j \in [y - \frac{n-1}{2}, y + \frac{n-1}{2}]\}.
    \end{split}
    \end{equation}
    \vspace{-3mm}

For the estimation of the anomaly map obtained in the previous iteration, we sort the probabilities (or errors) ascendingly, and take the indices corresponding to the smallest top $k^2$ elements, where $k^2$ is the number of trainable parameters in the convolution kernel and $k \leq n$,
\begin{equation}
    \mathcal{D}(x,y)=\operatorname{argsort} (\hat{\mathrm{\mathbf{M}}}_{\mathcal{N}(x,y)})[\,:k^2].
\end{equation}
\vspace{-3mm}

Finally, convolve the feature map with elements taken from the corresponding indices in $\mathcal{D}(x,y)$,
\begin{equation}
    \begin{aligned}
    \mathrm{\mathbf{F}}^{\prime}(x,y)&=\mathrm{\mathbf{F}}_{\mathcal{D}(x,y)}  \ast \mathrm{\mathbf{K}}  \\
    &= \sum_{i=1}^{k} \sum_{j=1}^{k}  \mathrm{\mathbf{F}}(d_i,d_j)   \cdot  \mathrm{\mathbf{K}}(i,j),
\end{aligned}
\end{equation}
where $ \mathrm{\mathbf{F}}^{\prime}$ is the feature obtained by AdaConv, and $\mathrm{\mathbf{K}}$ is the trainable kernel with size of $k \times k$. The uppooled features are reconstructed to the original image by performing a dot product with features extracted using AdaConv,
\begin{equation}
    \hat{\mathrm{\mathbf{X}}}^t=\mathrm{\mathbf{U}} \odot \mathrm{\mathbf{F}}^{\prime}.
\end{equation}
where $t$ represents the current iteration. $l_2$-norm is employed to calculate the anomaly score of pixel $p$, 
\begin{equation}
    \hat{\mathrm{\mathbf{M}}}_p^{t}=\left\|\hat{\mathrm{\mathbf{X}}}_p^{t}-\mathrm{\mathbf{X}}_p\right\|_{2},
\end{equation}
and the estimated detection map is used to guide the reconstruction process in the next iteration,
\begin{equation}
    \hat{\mathrm{\mathbf{X}}}^{t+1}=\mathcal{F}( \operatorname{SPP}(\mathrm{\mathbf{X}}), \hat{\mathrm{\mathbf{M}}}^{t}; \theta ).
\end{equation}
\vspace{-3mm}

\subsection{Design of the Regularization Term $\mathcal{R}$}

The regularization term imposes constraints on the anomaly map during the backward propagation of errors. However, determining the balance coefficient between the reconstruction and regularization terms remain challenging. We propose Online Background Pixel Mining (dubbed as OBPM) loss, which simultaneously achieves more efficient reconstruction and provides stronger constraints on anomalies. OBPM incorporates two key strategies: (1) For the reconstruction of the background, the more difficult the background is to reconstruct, the larger the gradient will be contributed. Gradient will be scaled exponentially with the reconstruction error. (2) For the regularization of anomalies, we enforce the disregard of gradients generated by potential anomalies. The two aspects ensure that the model focuses on reconstructing more complex background while avoiding the influence of anomalies that could distort the training process.

\subsubsection{Reconstructing Background}

Given the absolute background reconstruction error $x$, we desire that its backpropagation yields an exponentially scaled gradient,
\begin{equation}
    g(x) = e^{\beta x} + \alpha,
\end{equation}
here, the rate of exponential growth is determined by $\beta$, whereas $\alpha$ sets the minimum gradient. Thus, the reconstruction loss can be formulated as,
\begin{equation}
    l(x) = e^{\beta x} / {\beta} + \alpha x.
\end{equation}

\subsubsection{Regularizing Anomaly} 

The ideal solution is to not allow anomalies to contribute any gradients, \textit{i.e.}, discarding potential anomalies. Specifically, for superpixel $S_i$, the reconstruction error $\mathbf{e_i}$ will be firstly sorted ascendingly, 
\begin{equation}
\mathbf{e_i}^{\prime} = \operatorname{sort}(\mathbf{e_i}) = [e_1, e_2, \ldots, e_{|\mathcal{S}_i|}],
\end{equation}
where $e_1 \leq e_2 \leq \ldots \leq e_{|\mathcal{S}_i|}$. Since the basic assumption is that the anomalies has significantly larger errors than the background, we set the index with the largest error change as the boundary,
\begin{equation}
    q = \arg\max_{j} \{e_{j+1} - e_j \}, j = 1, 2, \ldots, |\mathcal{S}_i|-1,
\end{equation}
where $e_{j+1} - e_j$ represents the first order difference in sorted error, reflecting the magnitude of the error change. $q$ is the index where the error changes the most. Any error greater than $\mathbf{e_i}^{\prime}[q]$ will be ignored. Note that this will cause some background errors to be ignored, but since we provide exponential gradients, the remaining background can still provide enough gradients for network optimization. 

Combining the reconstruction loss, the OBPM loss of an error $x$ which belongs to $S_i$ is expressed as follows,
\begin{equation}
    \operatorname{OBPM}(x_{\in S_i})=
    \begin{cases}
        e^{\beta x} / \beta + \alpha x, & \text{ if } x \leq \mathbf{e_i}^{\prime}[q] \\
       0, & \text{ otherwise} .
      \end{cases}
\end{equation}

\begin{table*}[htbp]
  \centering
  \caption{AUC values of the 9 considered detectors on 12 datasets. \\ The best performance is shown in \textbf{bold} and the second best is \underline{underlined}.}
  \begin{tabular}{lccccccccc}
    \toprule
    Data$\setminus$Model & RXD~\cite{RXD} & CRD~\cite{CRD} & GAED~\cite{GAED} & MSNet~\cite{MSNet} & PDBSNet~\cite{PDBSNet} & PTA~\cite{PTA} & AutoAD~\cite{AutoAD} & RGAE~\cite{RGAE} & SuperAD (Ours) \bigstrut \\
    \hline
    Texas Coast & 0.9906 & 0.9910 & 0.9779 & 0.9946 & \underline{0.9950} & 0.6992 & 0.9938 & 0.9709 & \textbf{0.9982} \bigstrut[t] \\
    San Diego & 0.9089 & 0.8608 & 0.9866 & \underline{0.9907} & 0.9820 & 0.9683 & 0.9849 & 0.6991 & \textbf{0.9929} \bigstrut[t] \\
    HYDICE Urban & 0.9933 & 0.9975 & 0.9845 & 0.9993 & 0.9996 & 0.8659 & \textbf{0.9998} & 0.7064 & \underline{0.9993} \bigstrut[t] \\
    Pavia & 0.9537 & 0.9167 & 0.9362 & 0.9889 & \underline{0.9892} & 0.9061 & 0.9818 & 0.9053 & \textbf{0.9911} \bigstrut[t] \\
    ABU-Airport-1 & 0.8380 & 0.8481 & 0.8747 & \textbf{0.9582} & 0.9279 & 0.6504 & 0.9179 & 0.7773 & \underline{0.9418} \bigstrut[t] \\
    ABU-Airport-2 & 0.9502 & 0.7902 & 0.9049 & 0.9445 & \underline{0.9834} & 0.9663 & 0.9915 & 0.6698 & \textbf{0.9965} \bigstrut[t] \\
    ABU-Beach-1 & 0.9531 & \textbf{0.9960} & 0.9184 & 0.9580 & 0.9610 & 0.9303 & 0.9787 & 0.9470 & \underline{0.9844} \bigstrut[t] \\
    ABU-Beach-2 & 0.9106 & 0.9248 & 0.5444 & 0.9129 & \underline{0.9518} & 0.0960 & 0.9374 & 0.9049 & \textbf{0.9627} \bigstrut[t] \\
    ABU-Urban-1 & 0.9926 & 0.9394 & 0.9993 & \underline{0.9994} & 0.9994 & 0.8161 & 0.9960 & 0.9993 & \textbf{0.9994} \bigstrut[t] \\
    ABU-Urban-2 & 0.9501 & 0.9420 & 0.9591 & 0.9767 & \underline{0.9864} & 0.5087 & 0.9772 & 0.8249 & \textbf{0.9943} \bigstrut[t] \\
    ABU-Urban-3 & 0.9878 & 0.9719 & 0.9938 & \underline{0.9966} & 0.9966 & 0.4826 & 0.9908 & 0.9965 & \textbf{0.9970} \bigstrut[t] \\
    ABU-Urban-4 & 0.9685 & 0.8986 & 0.8104 & 0.9700 & \underline{0.9724} & 0.5192 & 0.9573 & 0.9651 & \textbf{0.9864} \bigstrut[t] \\
    \midrule
    Average & 0.9498 & 0.9231 & 0.9075 & 0.9741 & \underline{0.9787} & 0.7007 & 0.9756 & 0.8639 & \textbf{0.9870} \bigstrut[t] \\
    \bottomrule
  \end{tabular}%
  \label{tab:cmp-all-models}%
\end{table*}

\begin{figure*}[htbp]
  \centering
  \includegraphics[width=1\linewidth]{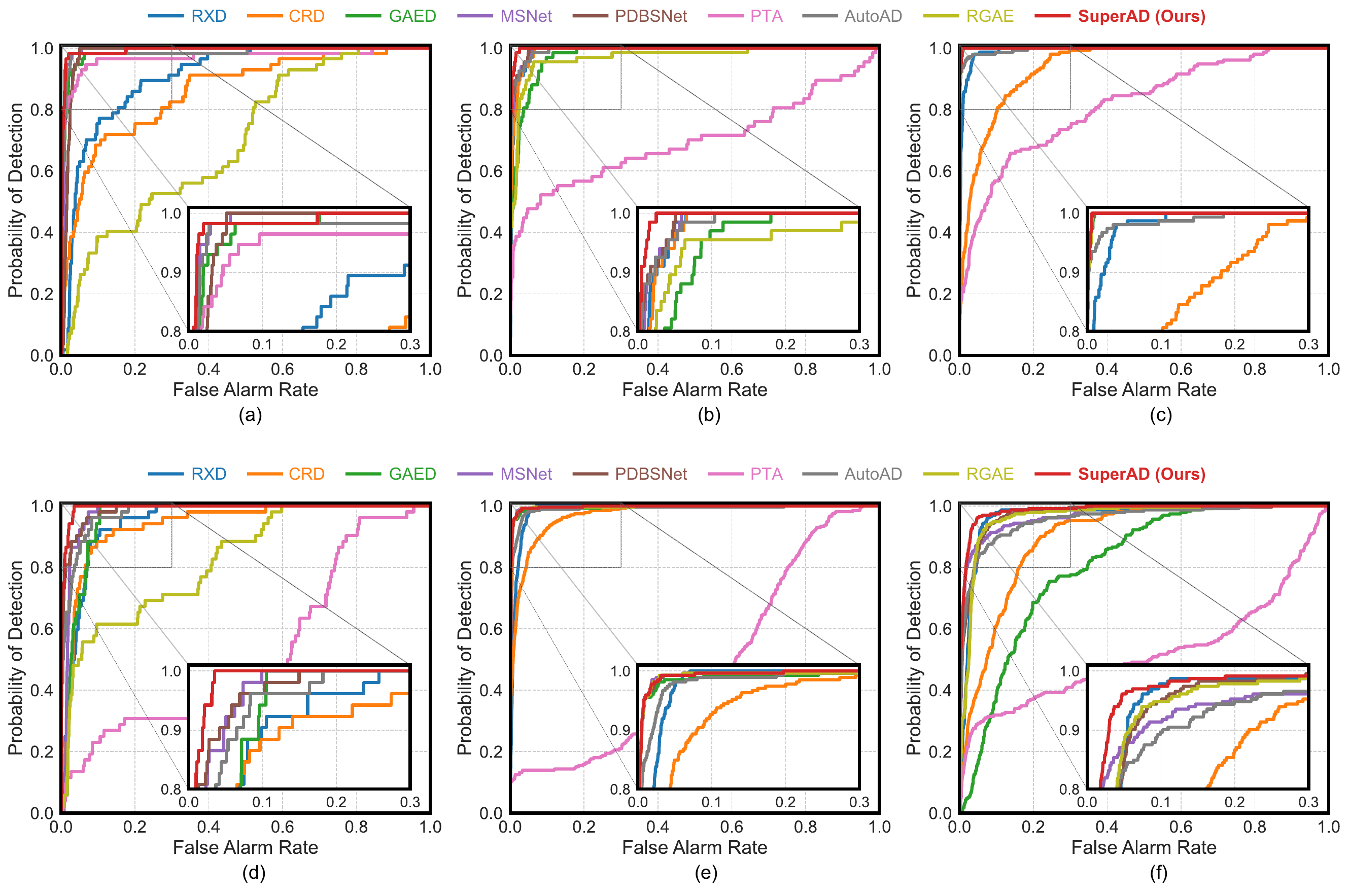}
  \caption{ROC curves of the 9 considered detectors on (a) San Diego, (b) Texas Coast, (c) ABU-Urban-1, (d) ABU-Urban-2, (e) ABU-Urban-3, and (f) ABU-Urban-4.}
  \label{fig:cmp-all-models-roc}
\end{figure*}

\begin{figure*}[htbp]
    \centering
    \includegraphics[width=1\linewidth]{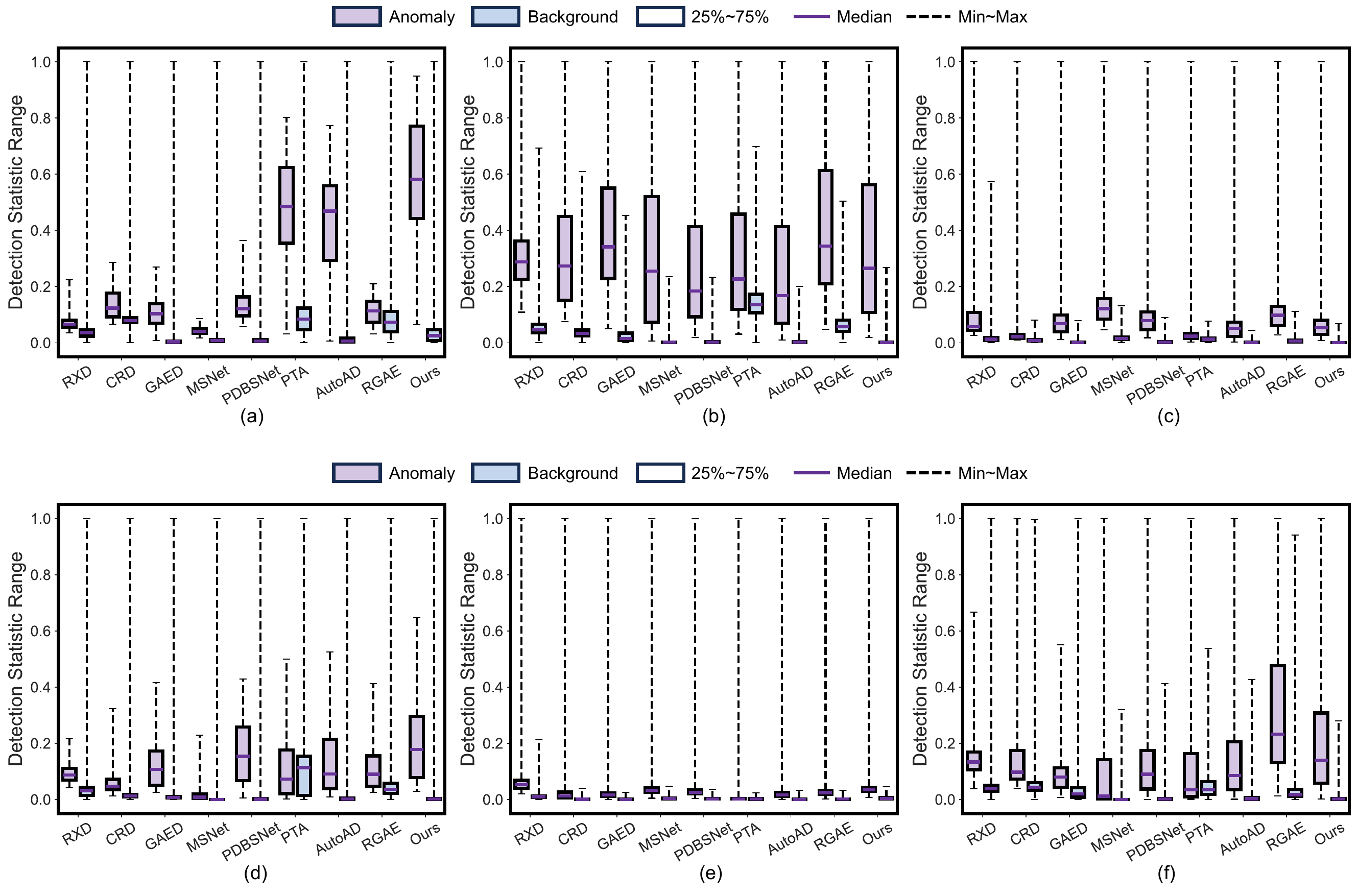}
    \caption{Separability maps of the 9 considered detectors on (a) San Diego, (b) Texas Coast, (c) ABU-Urban-1, (d) ABU-Urban-2, (e) ABU-Urban-3, and (f) ABU-Urban-4.}
    \label{fig:cmp-all-models-box}
  \end{figure*}

% $\hat{\mathrm{\mathbf{M}}} \in \mathbb{R}^{h\times w}$, $ \mathrm{\mathbf{F}} \in \mathbb{R}^{h\times w}$ 

\begin{figure*}[htbp]
  \centering
  \includegraphics[width=1\linewidth]{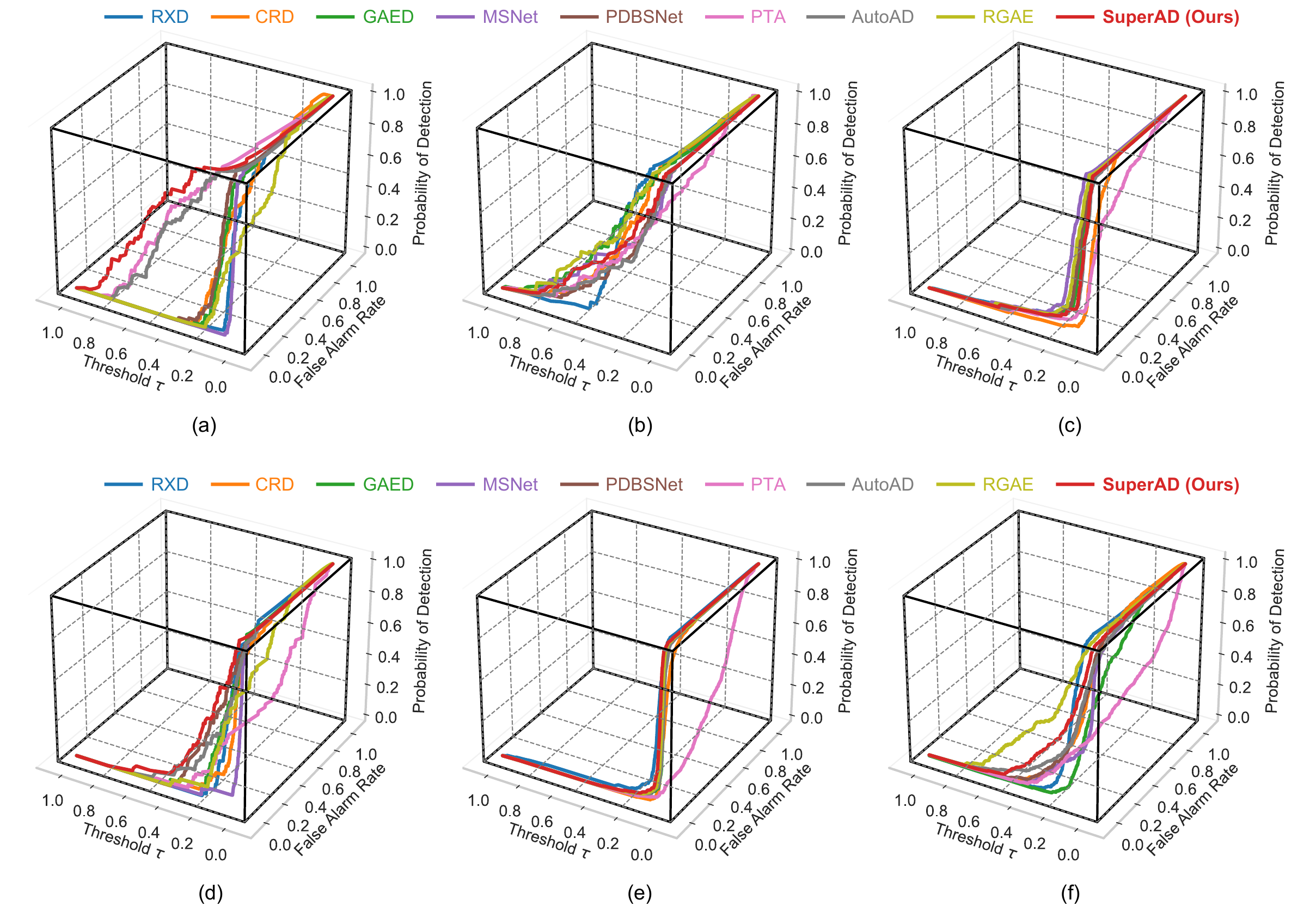}
  \caption{3D-ROC curves of the 9 considered detectors on (a) San Diego, (b) Texas Coast, (c) ABU-Urban-1, (d) ABU-Urban-2, (e) ABU-Urban-3, and (f) ABU-Urban-4.}
  \label{fig:cmp-all-models-3droc}
\end{figure*}

\section{Experimental Results}\label{Experimentss}

\subsection{Experimental Settings}

We evaluate the performance of our methods using seven widely recognized hyperspectral datasets: Texas Coast, San Diego, HYDICE Urban, Pavia, ABU-Airport, ABU-Beach and ABU-Urban. Eight commonly-recognized models including tranditional RXD~\cite{RXD} and CRD~\cite{CRD}, and self-supervised methods with diverse architectures including GAED~\cite{GAED}, MSNet~\cite{MSNet}, PDBSNet~\cite{PDBSNet}, PTA ~\cite{PTA}, AutoAD ~\cite{AutoAD}, and RGAE~\cite{RGAE}, were compared with the proposed methods.
The network architecture was implemented using PyTorch. All experiments were conducted on an NVIDIA GeForce RTX 2080 Ti with 11 GB of memory. Access the source code: \url{https://github.com/yc-cui/Super-AD}.

\begin{figure*}[htbp]
  \centering
  \includegraphics[width=1\linewidth]{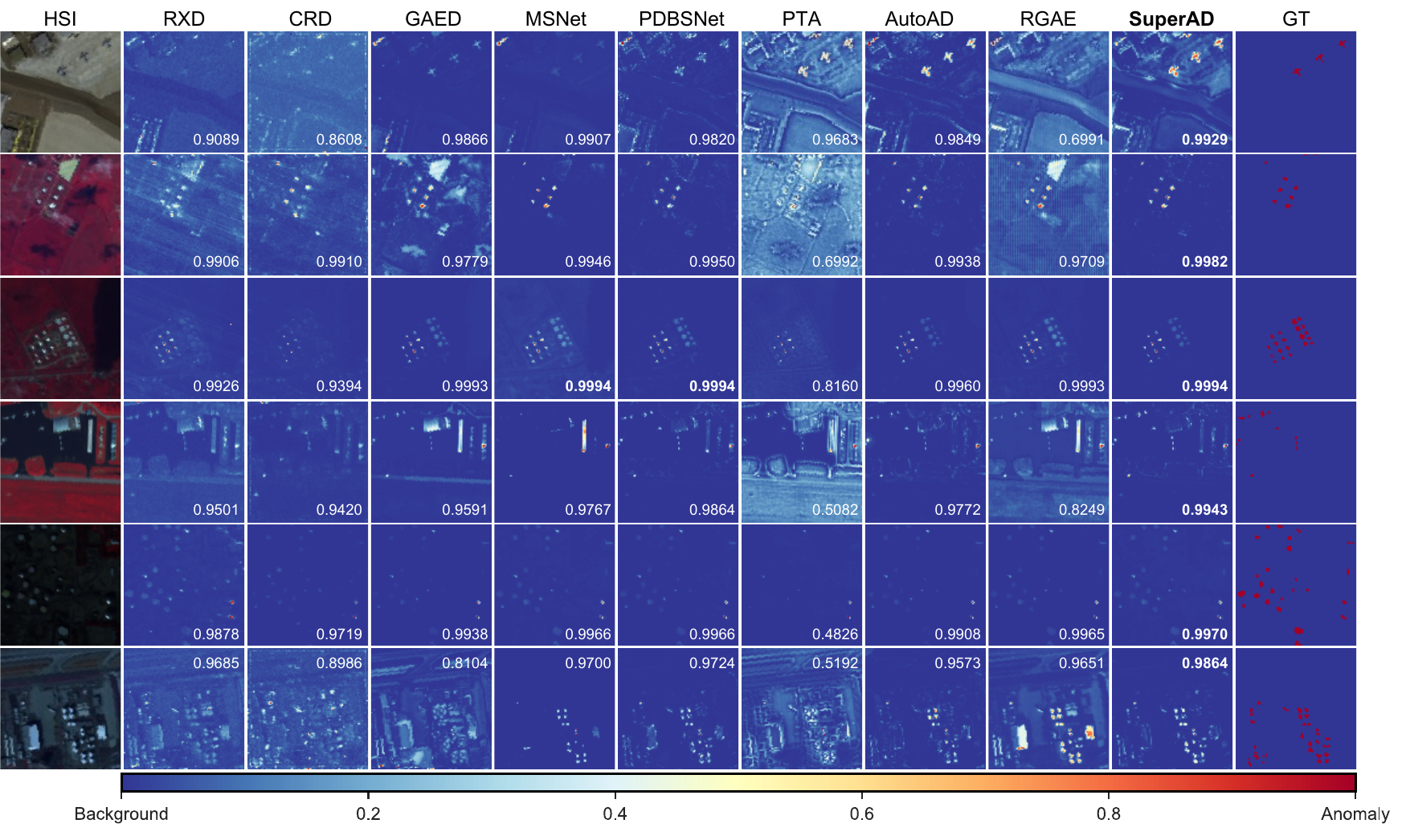}
  \caption{Visualization of detection results of the 9 considered detectors on (from top to bottom) San Diego, Texas Coast, ABU-Urban-1, ABU-Urban-2, ABU-Urban-3, and ABU-Urban-4. }
  \label{fig:cmp-all-models}
\end{figure*}

\subsection{Detection Performance Comparison}

\subsubsection{Quantitative Comparison}
As shown in Table \ref{tab:cmp-all-models}, our model obtained leading results in terms of AUC across the majority of the datasets. The performance differences between the datasets can be attributed to their distinct spectral-spatial features. On the San Diego dataset ($0.9982$ AUC), the anomalies (three airplanes with structural information) contrast spectrally with the background, and our spectral perturbation strategy significantly reduces the background interference. Although the performance is slightly degraded on the HYDICE Urban dataset ($0.9993$ vs. $0.9998$), where the vehicles and roofs are considered anomalies with $21$ anomalous pixels constituting about $0.26$\% of the entire image, the adaptive-weighted loss function in AutoAD~\cite{AutoAD} shows a localized advantage. But overall performance remains competitive. The comparative analysis of the receiver operating characteristic curve (ROC) presented in Fig.~\ref{fig:cmp-all-models-roc} and the separability maps shown in Fig.~\ref{fig:cmp-all-models-box} across all seven datasets further validates the superior ability of the proposed model to distinguish anomalies from background compared to other models. This consistent improvement highlights the effectiveness of our spectral perturbation strategy in enhancing feature discrimination.            

To further assess the effectiveness of the model, Fig.~\ref{fig:cmp-all-models-3droc} illustrates the comparative results of the three-dimensional ROC (3DROC) curves~\cite{chang2020effective,song20203}. 3DROC extends the evaluation dimensions of the traditional ROC by introducing a detection threshold $\tau$, whose key metric, the signal-to-noise ratio (SNPR), is calculated as:

\begin{equation}
\text{SNPR} = 10 \cdot \log_{10}\left(\frac{A_{\text{PD-}\tau}}{A_{\text{PF-}\tau}}\right)
\end{equation}

where $A_{\text{PD-}\tau}$ and $A_{\text{PF-}\tau}$ denote the area under the curve of the detection rate-threshold curve and false alarm rate-threshold curve, respectively. The experimental results demonstrate three key advantages of our method: First, in the PD-PF plane, our curve shows the closest proximity to the upper-right boundary with the largest $A_{\text{PD-PF}}$ area, indicating optimal detection rate-false-alarm rate trade-off across all threshold values. Second, the PD-$\tau$ projection reveals significantly larger $A_{\text{PD-}\tau}$ area compared to baseline methods, confirming superior target capture capability even under strict threshold constraints. Third, while our $A_{\text{PF-}\tau}$ area is not the absolute minimum, the maximized signal-to-noise ratio enables exceptional robustness in anomaly discrimination within complex interference environments. This analysis shows that the proposed framework effectively suppresses the false alarm accumulation effect while maintaining high detection sensitivity through the synergistic action of spectral perturbation and adaptive regularization, a feature that is difficult to achieve in conventional methods.

% Table generated by Excel2LaTeX from sheet 'Sheet5'
\begin{table}[htbp]
  \centering
  \caption{Comparison of model parameters, computational complexity, and training time. \textbf{Bold} indicates the best performance. $^\dag$ indicates MATLAB code.}
  \begin{tabular}{lccc}
    \toprule
    Model                   & \#Params (M) & MACs (G) & Avg. Time \bigstrut \\
    \hline
    RXD~\cite{RXD}$^\dag$   & -            & -        & 01m 26s \bigstrut[t] \\
    CRD~\cite{CRD}$^\dag$   & -            & -        & 01m 31s \bigstrut[t] \\
    GAED~\cite{GAED}$^\dag$ & -            & -        & 01m 33s \bigstrut[t] \\
    MSNet~\cite{MSNet}      & 0.553        & 5.511    & 05m 49s \bigstrut[t] \\
    PDBSNet~\cite{PDBSNet}  & 0.687        & 5.459    & 14m 09s \bigstrut[t] \\
    PTA~\cite{PTA}$^\dag$   & -            & -        & \textbf{00m 53s} \bigstrut[t]    \\
    AutoAD~\cite{AutoAD}    & 3.249        & 5.979    & 06m 04s \bigstrut[t]  \\
    RGAE~\cite{RGAE}$^\dag$ & -            & -        & 02m 01s \bigstrut[t]  \\
    SuperAD (Ours)          & \textbf{0.241}        & \textbf{2.220}    & 05m 50s \bigstrut[t] \\
    \bottomrule
  \end{tabular}%
  \label{tab:cmp-efficiency}%
\end{table}%

\subsubsection{Visual Comparison}
Fig.~\ref{fig:cmp-all-models} illustrates the anomaly detection maps for the San Diego, Texas Coast, and ABU-Urban datasets. Among all the evaluated models, only MSNet~\cite{MSNet} and AutoAD~\cite{AutoAD} demonstrate competitive performance to our approach. While MSNet~\cite{MSNet} yields impressive results on the Texas Coast dataset, it struggles to identify anomalous pixels within the San Diego dataset. This may be due to the fact that the network is not designed for use in urban zones with complex background information. For AutoAD~\cite{AutoAD}, although it also exhibited strong performance, our model assigns higher probabilities to anomalous points compared to AutoAD~\cite{AutoAD}. This clearly indicates that the proposed model effectively differentiates anomalies from the background, highlighting the efficacy of our method. In practice, it is possible to consistently assign higher probabilities to true anomalies while minimizing false positives. The advanced of our model on different datasets demonstrates that the approach we employ (perturbation, reconstruction, and regularization) is suitable and robust for handling the complexity and variability present in different hyperspectral imaging scenarios. This consistent performance across datasets enhances the robustness and reliability of our approach in hyperspectral anomaly detection.

\subsubsection{Comparison of Model Efficiency}
The model complexity presented in Table~\ref{tab:cmp-efficiency} demonstrates the advantages of our proposed model according to the number of parameters and computational complexity. With only $0.241$ million parameters, our architecture achieves a significant reduction in model complexity compared to existing approaches such as AutoAD~\cite{AutoAD}. The computational complexity, measured in multiply-accumulate operations (MACs), further underscores the efficiency of our approach. The MACs was calculated for an input tensor of size $(1, 204, 100, 100)$. Our model requires less than half the computational resources of comparable methods. The training time reveals that our model maintains competitive efficiency, achieving performance comparable to MSNet~\cite{MSNet} while being significantly faster than PDBSNet~\cite{PDBSNet}.

\subsection{Ablation Study and Parameter Analysis}

\begin{table}[t]
  \centering
  \caption{Ablation of perturbation operation SPP.}
  \begin{tabular}{rccccc}
    \toprule
            & Coast           & San Diego       & HYDICE Urban          & Pavia           & Average \bigstrut[t]      \\
    \midrule
    w/o SPP & 0.9938          & 0.9905          & 0.9960          & 0.9847          & 0.9913 \bigstrut          \\
    w/ SPP  & \textbf{0.9982} & \textbf{0.9929} & \textbf{0.9993} & \textbf{0.9911} & \textbf{0.9954} \bigstrut \\
    \bottomrule
  \end{tabular}%
  \label{tab:abl-SPP}%
\end{table}%

\subsubsection{Perturbation Operation SPP}\label{sec:spp}
An ablation study was conducted to evaluate the contribution of the superpixel pooling and uppooling mechanism, as shown in Table \ref{tab:abl-SPP}. The results clearly demonstrate the significant impact of SPP on the model's performance, with a noticeable increase in AUC scores when SPP is incorporated. This indicates that SPP plays a crucial role in mitigating the IMP by encapsulating anomalous pixels within background-dominated blocks, thereby preventing their influence on the reconstruction process.

\begin{figure}[htbp]
  \centering
  {\includegraphics[width=0.6\linewidth]{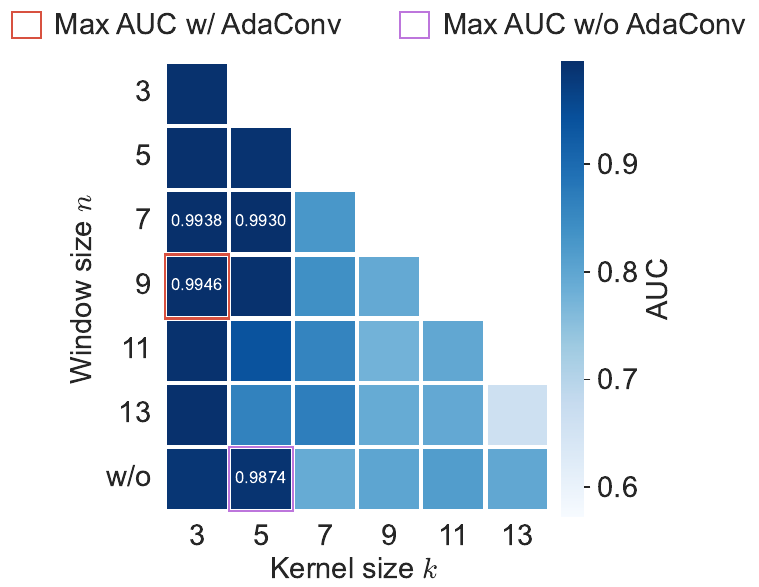}}
  \caption{Ablation studies and parameter analysis conducted on San Diego dataset for AdaConv.}
  \label{fig:abl-nk}
\end{figure}

\subsubsection{Reconstruction Function AdaConv}
Fig.~\ref{fig:abl-nk} present the results of ablation studies and parameter analysis on the reconstruction function AdaConv. Optimal performance is achieved with a window size of $n=9$ and a kernel size of $k=3$, yielding an AUC of $0.9946$. We noticed AdaConv exhibits sensitivity to large kernels, such as \{7, 9, 11, 13\}, possibly due to the incorporation of irrelevant information by distant pixels, which diminishes the local correlation with the center pixel. As shown in the last row of Fig.~\ref{fig:abl-nk}, without AdaConv, the optimal AUC is $0.9874$. This indicates that AdaConv effectively targets non-anomalous pixels, enhancing the model's ability to reconstruct the background while disregarding anomalies, thereby preventing IMP and achieving superior results.

\begin{figure}[htbp]
  \centering
  {\includegraphics[width=0.8\linewidth]{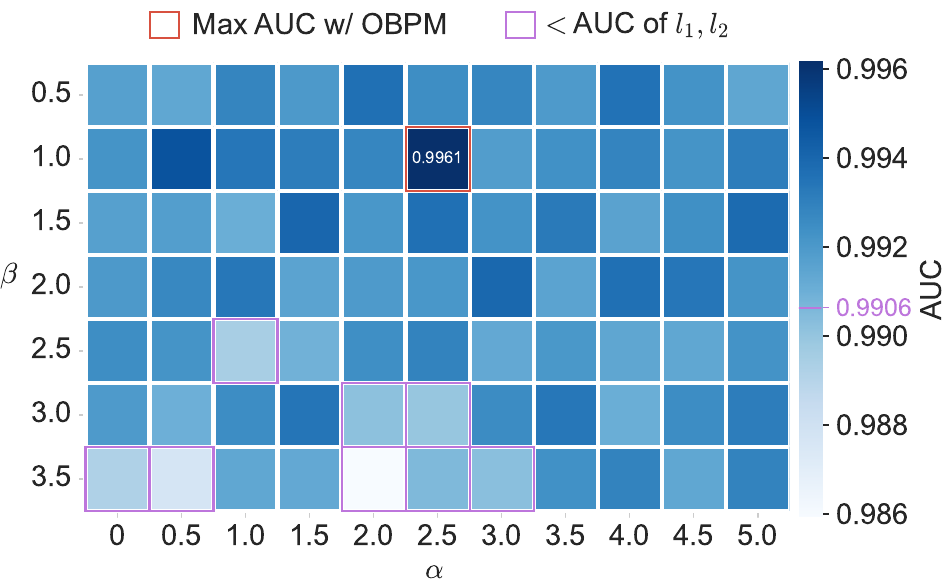}}
  \caption{Ablation studies and parameter analysis conducted on San Diego dataset for OBPM.}
  \label{fig:abl-ab}
\end{figure}

\begin{figure*}[htbp]
  \centering
  \includegraphics[width=1\linewidth]{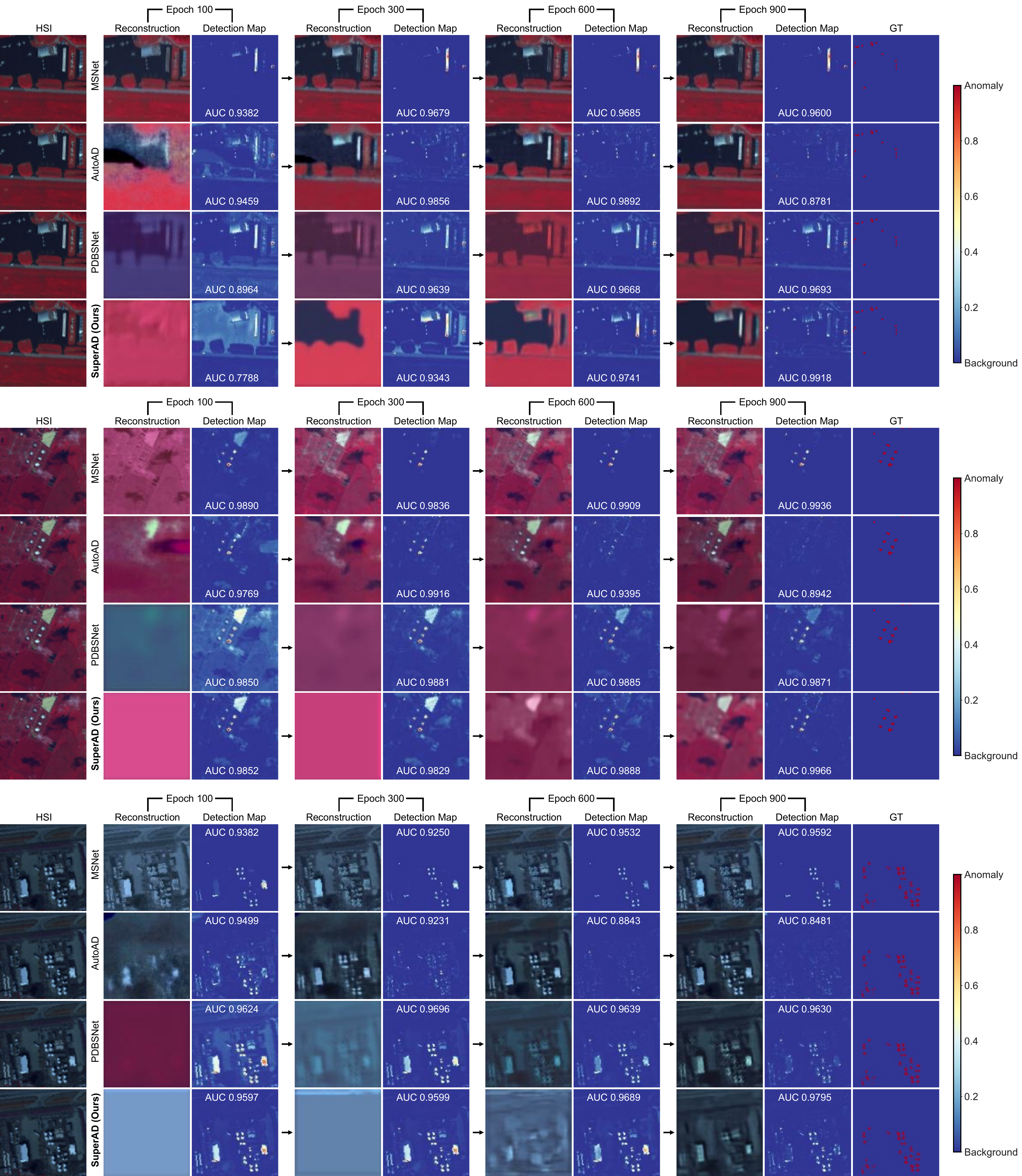}
  \caption{Visualization of the reconstruction process of different detectors.}
  \label{fig:imp-viz}
\end{figure*}

\subsubsection{Regularization Term OBPM}
Fig.~\ref{fig:abl-ab} illustrates the effectiveness of the OBPM. Our method achieved an optimal AUC of $0.9961$ compared to $0.9906$ achieved by the commonly used $l_1$ or $l_2$ loss. The OBPM strategy is shown effective on reconstructing complex backgrounds while disregarding potential anomalies. The parameter sensitivity analysis reveals that the proposed OBPM performs consistently well within a range for $\beta \in [0.5, 2]$ and $\alpha \in [0, 5]$, indicating its robustness and stability across different parameter settings.

The parameter sensitivity analysis of superpixel segments, as illustrated in Fig.~\ref{fig:ana_segs}, reveals the remarkable robustness of our SPP mechanism across varying numbers of superpixel blocks. The experimental results demonstrate consistent detection performance when the number of superpixels ranges from $10$ to $900$. This stability can be attributed to the inherent property of SPP to effectively eliminate anomalous points regardless of the number of superpixel segmentation, as further validated by the detailed visualization results presented in Sec.~\ref{sec:spp} in Fig.~\ref{fig:viz-spp}. The ablation studies in Fig.~\ref{fig:ana_segs} further substantiate the essential role of both AdaConv and OBPM in our framework. The significant performance degradation observed when either component is removed underscores their critical contributions to the model's effectiveness. These findings collectively reinforce the validity and robustness of our proposed approach in addressing the IMP in self-supervised HAD.

\subsection{Visualization Analysis}

\subsubsection{Visualization of Reconstruction Results}
To better illustrate the superiority of our proposed model in addressing the IMP, we present a step-by-step visualization of the reconstruction process. We compare our model with three other reconstruction-based models (AutoAD~\cite{AutoAD}, PDBSNet~\cite{PDBSNet}, and MSNet~\cite{MSNet} ), as shown in Fig.~\ref{fig:imp-viz}. The visualization results clearly demonstrate that our model effectively mitigates the IMP throughout the training process. Specifically, as the number of epochs increases, our model consistently fails to reconstruct anomalous pixels, while successfully reconstructing the background. In contrast, AutoAD~\cite{AutoAD} tends to reconstruct the entire input image, including anomalies, leading to degraded detection performance. Similarly, PDBSNet~\cite{PDBSNet} and MSNet~\cite{MSNet} show varying degrees of anomaly reconstruction, particularly in later epochs, which compromises their ability to distinguish anomalies from the background. This comparative analysis underscores the effectiveness of our approach in preventing the reconstruction of anomalous spectra, thereby maintaining robust anomaly detection capabilities throughout the training process.

\begin{figure}[t]
  \centering
  \includegraphics[width=1\linewidth]{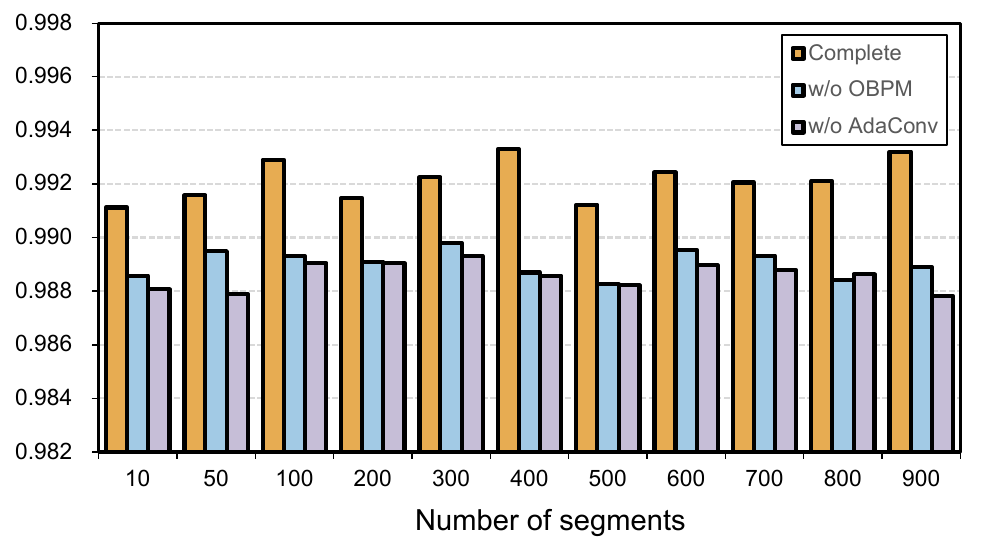}
  \caption{Detection performance of SuperAD with different numbers of superpixel segments.}
  \label{fig:ana_segs}
\end{figure}

\begin{figure}[htbp]
  \centering
  \includegraphics[width=1\linewidth]{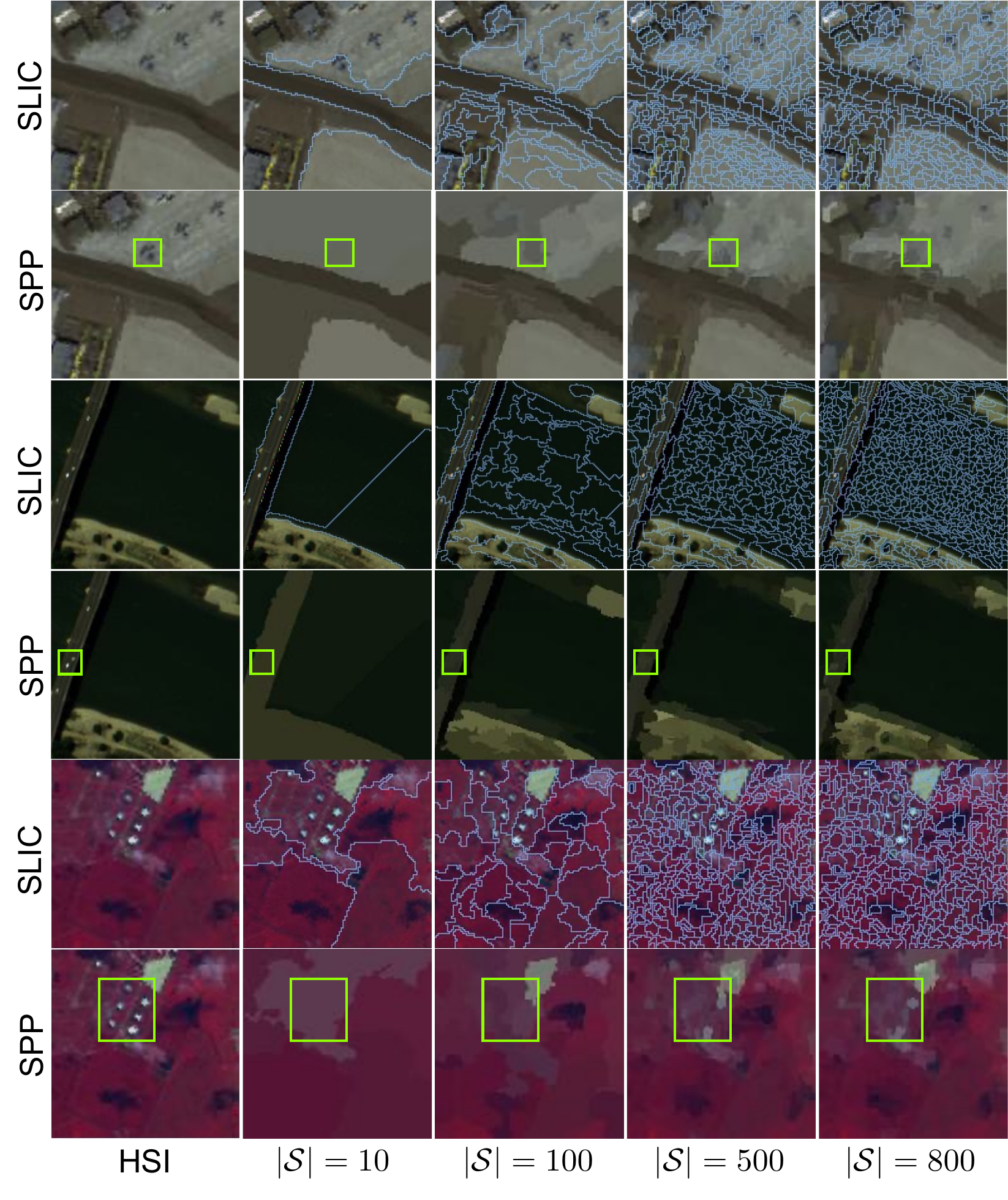}
  \caption{Visualization of superpixel pooling with different numbers of superpixels.}
  \label{fig:viz-spp}
\end{figure}

\subsubsection{Visualization of Superpixel Pooling}
Fig.~\ref{fig:viz-spp} presents a comprehensive visualization of our proposed superpixel pooling mechanism with varying numbers of superpixels, \textit{i.e.,} $|\mathcal{S}|=\{10, 100, 500, 800\}$. The highlighted regions indicate the locations of anomalous points. As the number of superpixels increases, the SPP demonstrates enhanced capability in preserving both spectral and spatial information. This is evident from the progressively detailed representation of the image structure across different superpixel counts. However, despite this increased information retention, the SPP consistently maintains its fundamental characteristic of effectively eliminating anomalous points from the reconstruction process. 

The visualization reveals that regardless of the superpixel count, the anomalous regions remain effectively suppressed in the supperpixeled images. As discussed in Sec.~\ref{sec:spp}, the stability of the SPP is also corroborated by the experimental results shown in Fig.~\ref{fig:ana_segs}, where the detection performance remains relatively stable despite variations in the number of superpixels. This parameter insensitivity is particularly advantageous in practical applications, as it reduces the need for meticulous parameter tuning while maintaining high detection accuracy. Combining the visualization results in Fig.~\ref{fig:viz-spp} and the experimental analysis in Fig.~\ref{fig:ana_segs}, our SPP approach demonstrates remarkable robustness in mitigating the influence of anomalies on the reconstruction process and thus prevents the IMP. The effectiveness of SPP can be attributed to its unique design: by encapsulating anomalous pixels within background-dominated blocks through the average pooling strategy, it inherently suppresses the influence of anomalies while preserving the essential characteristics of the background. This dual capability of information preservation and anomaly suppression contributes significantly to the overall superiority of our proposed method.

\begin{figure}[htbp]
  \centering
  \includegraphics[width=1\linewidth]{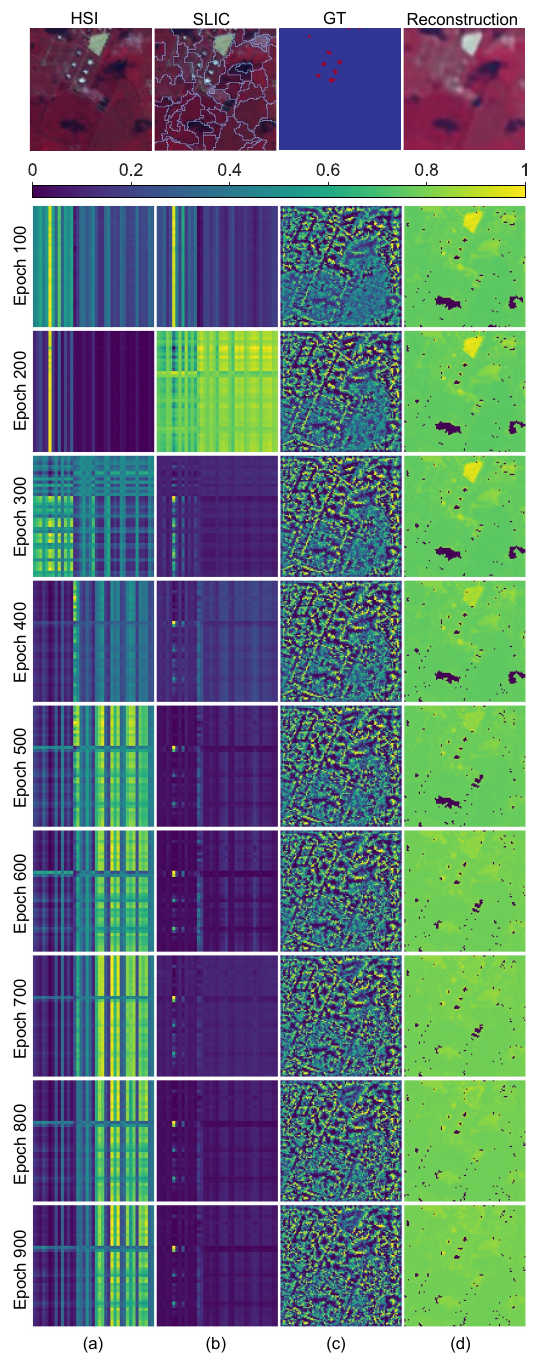}
  \caption{Visualization of training process: (a) and (b) SPP self-attention score matrices visualization. (c) AdaConv pixel utilization visualization. (d) OBPM  gradients visualization.}
  \label{fig:inter-viz}
\end{figure}

To further visualize the interaction mechanisms between different superpixel blocks during the reconstruction process, we present the attention score matrices from the first and final layers of the self-attention mechanism in SuperAD, as illustrated in Fig.~\ref{fig:inter-viz}\textcolor{blue}{(a)} and Fig.~\ref{fig:inter-viz}\textcolor{blue}{(b)}. The attention maps demonstrate that the inter-block relationships stabilize as the iteration progresses, with minimal observable changes in the attention patterns during later stages. Notably, the attention maps exhibit a vertical banded structure, indicating that the relative importance weights remain consistent for most blocks during the reconstruction process. However, a small subset of blocks, represented by the significantly brighter rows in Fig.~\ref{fig:inter-viz}\textcolor{blue}{(b)}, demonstrates unique reconstruction patterns. These blocks exhibit distinct attention distributions compared to the majority of blocks. This phenomenon suggests that while most blocks maintain consistent reconstruction dependencies, certain regions require specialized attention patterns to achieve accurate reconstruction.

\subsubsection{Visualization of AdaConv}
Fig.~\ref{fig:inter-viz}\textcolor{blue}{(c)} presents the visualization results of AdaConv, where the kernel size and window size are configured as $3$ and $5$, respectively. The visualization is generated by accumulating the number of pixels selected by the convolution kernel during the reconstruction process, normalized to the range of $[0,1]$. A value of $0$ indicates that the corresponding pixel was never utilized throughout the reconstruction process. The visualization reveals several critical insights into AdaConv. Notably, anomalous pixels consistently demonstrate zero utilization across all epochs, indicating their complete exclusion from the reconstruction process. This selective exclusion mechanism effectively prevents the influence of anomalies on the reconstruction outcome, thereby mitigating the IMP. However, this selective process also results in the partial sacrifice of background pixel utilization. The evolution of pixel utilization patterns across different training epochs provides further insights into AdaConv. During the initial training phase (\textit{e.g.,} at epoch $100$), the utilization patterns in smooth regions of the hyperspectral image exhibit relatively similar characteristics, suggesting a more generalized approach to background reconstruction. As the training progresses to later stages (\textit{e.g.,} at epoch $900$), we observed more refined and differentiated pixel utilization patterns. As iteration gains, regions with complex spectral variations tend to exhibit higher and more diverse utilization values compared to homogeneous areas. This adaptive behavior contributes to the model's ability to accurately reconstruct intricate background patterns while maintaining its robustness against anomaly contamination. The observed characteristics of AdaConv, including its selective pixel utilization, adaptive reconstruction strategy, and progressive refinement during training, collectively contribute to its effectiveness in addressing the IMP in self-supervised HAD.

\subsubsection{Visualization of OBPM}
Fig.~\ref{fig:inter-viz}\textcolor{blue}{(d)} illustrates the visualization of the proposed online background pixel mining, where gradient values are normalized within the range of $[0,1]$. Pixels with zero values indicate spatial locations that do not contribute gradients to the backpropagation process. The visualization demonstrates a progressive learning pattern: During initial iterations, certain background regions are temporarily excluded. However, as training progresses, these background gradients are progressively reintegrated. This adaptive mechanism effectively addresses the IMP by dynamically incorporating background information while consistently filtering out anomalies. The final results show that the majority of background pixels are successfully incorporated into the model optimization, with only a minimal fraction being erroneously excluded as potential anomalies. Notably, anomalous pixels maintain zero gradients throughout the entire training process, confirming the effectiveness of OBPM in preventing anomaly reconstruction. These findings highlight OBPM's capability that dynamically adjusts background gradient utilization while strictly maintaining anomaly exclusion, thereby significantly enhancing the discriminative ability of SuperAD to distinguish normal and anomalous pixels.

\section{Discussion and Conclusion}\label{Discussion and Conclusion}

This paper presents a novel approach to address the identity mapping problem in self-supervised HAD, which is grounded in a unified framework that encompasses three critical aspects: perturbation, reconstruction, and regularization. Through extensive experiments on various hyperspectral datasets, we have demonstrated the effectiveness of our proposed solutions, including superpixel pooling and uppooling, error-adaptive convolution, and online background pixel mining. Our work presents a significant step forward in the field of self-supervised HAD, offering a robust and effective approach to tackle the challenges posed by the IMP. It is hoped that this paper will provide valuable insights and inspire further research for self-supervised HAD.

Building on the proposed framework, future research could explore the proposed three directions to further enhance self-supervised hyperspectral anomaly detection. For instance, developing adaptive perturbation strategies that dynamically adjust superpixel segmentation scales or integrating spectral-spatial masking to enhance anomaly suppression. Additionally, advanced regularization techniques, such as hierarchical background modeling or uncertainty-aware loss functions, could strengthen robustness against false positives.

%%%%%%%%%% rev
% \nolinenumbers
\bibliographystyle{IEEEtran}
\bibColoredItems{revisioncolor}{10285392,10330008,TV,FE-HPM,Pan-Mamba,Lambda-PNN,duancvpr2024,ran2024knlconv} %新增ref，颜色设置为revisioncolor
\bibliography{egbib}

\end{document}